\documentclass[12pt]{iopart}

\usepackage{iopams} 
\usepackage{graphicx}
\usepackage{siunitx}
\usepackage[numbers]{natbib}
\usepackage{xcolor}
\usepackage{caption}
\usepackage{subcaption}

\begin{document}

\title{Flow turning effect and laminar control by the 3D curvature of leading edge serrations from owl wing }

\author{Muthukumar Muthuramalingam$^{1}$, Edward Talboys$^1$, Hermann Wagner$^2$, Christoph Bruecker$^1$}

\address{$^1$ City, University of London, Northampton Square, London, EC1V 0HB, UK}
\address{$^2$ RWTH Aachen University, Templergraben 55, 52062 Aachen, Germany}
\ead{muthukumar.muthuramalingam@city.ac.uk}
\vspace{10pt}
\begin{indented}
\item[]July 2020
\end{indented}

\begin{abstract}
This work describes a novel mechanism of laminar flow control of a backward swept wing with a comb-like leading edge device. It is inspired by the leading-edge comb on owl feathers and the special design of its barbs, resembling a cascade of complex 3D-curved thin finlets. 
The details of the geometry of the barbs from an owl feather were used to design a generic model of the comb for experimental and numerical flow studies with the comb attached to the leading edge of a flat plate.
Examination was carried out at different sweep angles, because life animal clearly show the backward sweep of the wing during gliding and flapping. 
The results demonstrate a flow turning effect in the boundary layer inboards, which extends along the chord over distances of multiples of the barb lengths.
The inboard flow-turning effect described here, thus, counter-acts the outboard directed cross-span flow typically appearing for backward swept wings. 
From recent theoretical studies on a swept wing, such a way of turning the flow in the boundary layer is known to attenuate crossflow instabilities and delay transition. A comparison of the comb-induced cross-span velocity profiles with those proven to delay transition in theory shows excellent agreement, which supports the laminar flow control hypothesis. 
Thus, the observed effect is expected to delay transition in owl flight, contributing to a more silent flight. 
\end{abstract}

\vspace{2pc}
\noindent{\it Keywords}: swept wing, leading-edge comb, laminar flow control

\submitto{\BB}


\section{Introduction}
One of the remaining puzzles in the silent flight of owls is the function of the serrated leading edge. This `comb-like' structure is more developed in nocturnal than diurnal owl species \citep{Weger2016}, suggesting that the leading-edge comb must have some benefit for hunting in the night.
Indeed it was suggested early on \citep{graham1934, lilley1998} that the serrations are one of the adaptations found in owls that underlie silent flights, where the owl needs to be as quiet as possible when hunting nocturnally.  
Acoustic measurements by \citet{neuhaus1973} and \citet{geyer2020} support this suggestion, although the effect was marginal for low angles of attack, the situation being relevant for the gliding phase persisting up to the final phase of direct attack of the prey. Alternative suggestions for their function were focusing on a possible aerodynamic benefit of a serrated leading edge \citep{Hertel1963,Kroeger1972,klaen2010, winzen2014, wagner2017,Rao2017,ikeda2018,wei2020}, summarized in the most recent review given in 2020 by \citet{Jaworski2020}.

An early contribution interpreted the leading edge comb as a tripping device, which triggers the boundary layer to turbulent transition, keeping the flow over the aerofoil attached \citep{Hertel1963}. 
This, however, would cause some extra turbulent noise, which is not observed \citep{geyer2020}. \citet{Kroeger1972} presented a comprehensive study of the flow around the leading edge of an owl wing.
Using wool tufts, these authors showed a spanwise flow behind the comb, which they interpreted as a way to prevent flow separation. 
Acoustic measurements by these authors, however, showed no direct influence of the presence of the comb. 
It was only at high angles of attack that a difference of about 3~dB was noticeable. 
This result was later confirmed by \citet{geyer2020} using acoustic 2D sound maps. 
These authors could show that the sources of higher noise levels for high angles of attack stem from the wing tip. \citet{Jaworski2020} speculated that the leading edge comb may play a role in reducing spanwise flow variations due to separation at high angles of attack, thereby reducing the strength of the tip vortex and the associated tip noise \citep{Jaworski2020}. 
If so, it would, however, not be relevant for the gliding phase.    

In a similar way, aerodynamic performance measurements on wings with serrated leading edge show  benefits mostly with increasing angle of attack, again not much relevant for the gliding phase. \citet{Rao2017} showed that planar leading-edge serrations can passively control the laminar-to-turbulent transition over the upper wing surface. Each of the serrations generates a vortex pair, which stabilizes the flow similar as vortex generators do.  \citet{wei2020} applied such serrations on the wing of a propeller to shift the location of laminar-to-turbulent transition on the suction side. \citet{ikeda2018} investigated different length of the serrations to find the optimum of lift-to-drag ratio at angles of attack $< 15^\circ$.

A remaining contribution to noise reduction at gliding flight conditions may be the influence of the comb on leading-edge noise from incoming vortices and unsteady flow components present in the air environment.
To test this hypothesis, researchers investigated the noise emission of wings in an anechoic wind tunnel with unsteady inflow conditions generated by an upstream inserted turbulence grid \cite{geyer2017-LEC}. 
The results showed that serrations can attenuate unsteady flow effects caused by oncoming vortices and turbulence. Similar results were found from LES simulations of serrations in turbulent inflow conditions \citep{chaitanya}. These findings agree with measurements on noise emission of stationary aerofoils where artificial serrations led to a lower noise radiation in unsteady flow \citep{geyer2017-LEC, Narayanan2015}. 

Herein, we introduce a novel hypothesis which is related to the influence of serrations on swept wing aerodynamics. First, data of owls in gliding flight clearly demonstrate that the wing's leading edge is swept backward, about 10--20$^\circ$, see Figure \ref{fig: FeatherScan} (adapted from snapshots of the movie produced in \citet{Durston2019} for a gliding American barn owl). Second, the serrations in nature are curved in a complex 3D shape protruding out of the plane of the wing \citep{Bachmann2011}. All of this may influence the flow over the wing and probably - by the complex coupling between flow and sound generation - it may influence also the overall noise emission. For swept wings it is known that a backward sweep can introduce considerable cross-flow instabilities, which trigger transition \citep{serpieri, Radeztsky, Edward}, invoking the substantially drag-increasing turbulent boundary-layer state \citep{Wassermann2002}. To overcome this drag penalty, flow control methods such as suction \citep{Kloker2008} and plasma actuators \citep{Dorr2015} have been developed to attenuate the instabilities. The present work demonstrates, that a similar effect may be achieved in a passive way by using a comb-like leading-edge structure with 3D curved finlets, inspired from the geometry of serrations on the owl wing. We show in the following that the serrations cause a change in flow direction near the wall (flow turning) at sweep angles observed in nature, thereby delaying transition and contributing to a more silent flight.

\section{Methods}
\subsection{Coordinate System of the wing}\label{sec: Method-Coordinates}
\begin{figure}[h]
\centering\includegraphics[scale = 0.4]{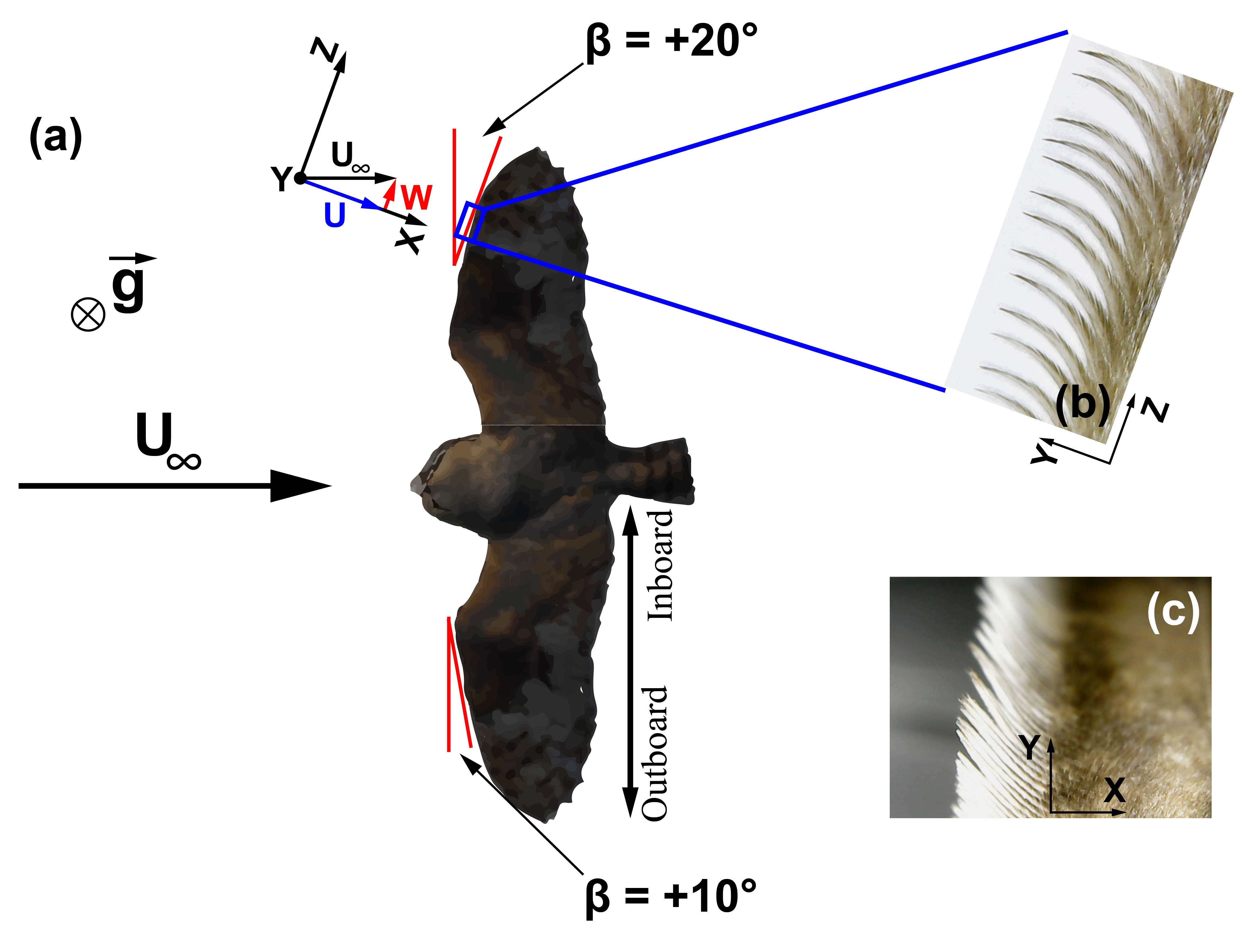}
\caption{Gliding owl and leading edge serrations. a) Top view of an owl in gliding flight, illustrating the backward sweep of the wing. The situation is shown in a body-fixed observer situation with wind coming from left at a velocity ($U_\infty$). The wing portion at mid span has an effective positive sweep angle of $\beta$ $\approx$ 10\si{\degree}, increasing to $\beta$ $\approx$ 20\si{\degree} further towards 3/4 span. The picture of the owl is reproduced/adapted from the video published in \citep{Durston2019} with permission from Journal of Experimental Biology, reference \citep{Durston2019} with DOI: 10.1242/jeb.185488. Inset b) pointed picture of leading edge comb in back view with flow coming out of the paper plane; inset c) pointed picture of side view of the serrations with flow coming from left .} \label{fig: FeatherScan}
\end{figure}

The world coordinate system of the flying body is typically defined in relation to the body axes and the direction of the flight path. 
Herein, we define (in capital letters) another Cartesian coordinate system which is fixed with the wing and oriented with the leading edge, see Fig. 1. The positive X-axis points in chordwise direction, the positive Y-axis vertically upwards, and the positive Z-axis is aligned with the leading edge of the wing (Fig. \ref{fig: FeatherScan}).The same coordinate system was used to describe the morphology of the leading edge comb of the owl feather in nature and for the model data, see Table \ref{table: Design Spec}. As a research platform of swept wing instabilities, often a flat swept plate was chosen for better control of the boundary conditions and access for the measurement methods \citep{Abegg}. Therefore the flat swept plate is considered as a generic testing model in the relevant community and is used herein for the same reasons of better access for CFD and experimental studies. Additional wing curvature effects on laminar-turbulent transition can be simulated by imposing either a negative or a positive pressure gradient on the potential flow outside \citep{Abegg}. 

\subsection{Generation of the generic comb model}
As may be seen in Fig. 1b, the feather that forms the leading edge has an outer vane with separated, filamentous barb endings. 
These barb endings are the serrations \citep{Bachmann2011}. 
Many parallel serrations form a leading edge comb-like structure. 
Each single serration has a complex shape with strong curvature in two major planes of the feather, the frontal Y-Z plane and the cross-sectional X-Y plane \citep{Bachmann2011}. 

A generic model of the leading edge comb was built based on data available in \citep{Bachmann2011}. 
The model consists of a series of barbs.
Each barb starts with the root and ends with the tip.
While the roots of the serrations are connected to each other, the tips are separated.
In the following we first describe the properties of the single barbs in more detail, before we explain how the barbs are aligned to form a leading-edge comb.

\begin{figure}[h]
\centering\includegraphics[scale = 0.3]{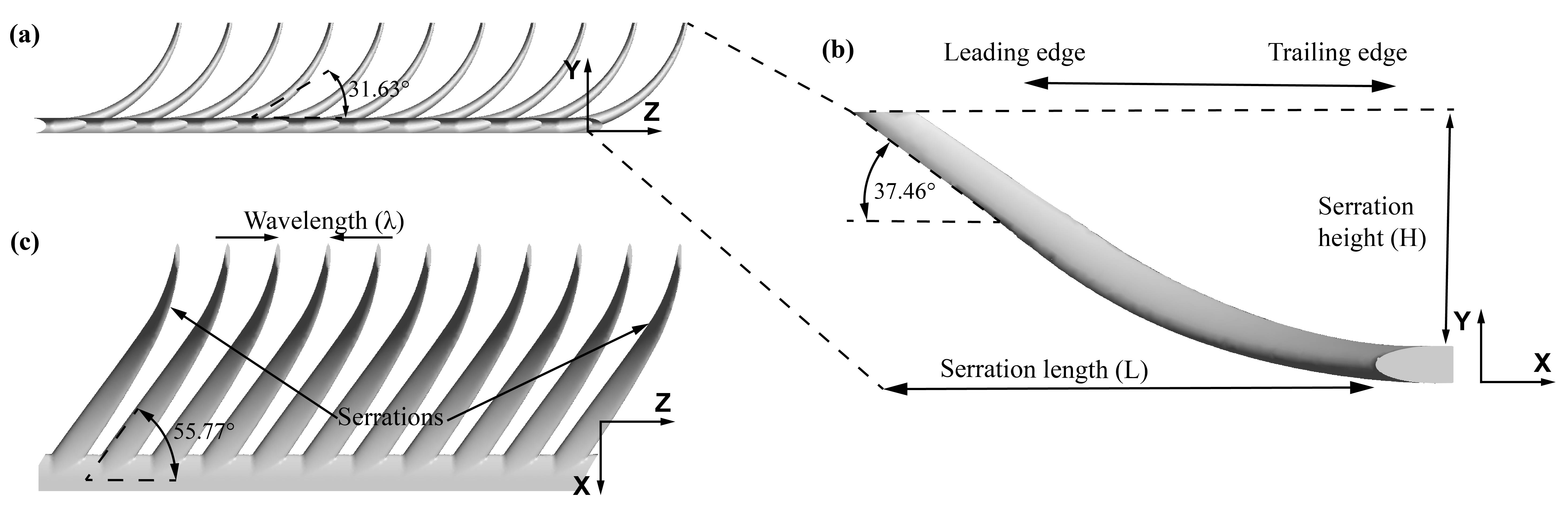}
\caption{Orientation of the reconstructed serrated leading edge. a) back-view of the comb, locking from the back over the feather onto the outstanding barbs of the right wing, compare also Fig. 1b. b) Side view on a single barb in enlarged scale c) top-view of the comb in the feather plane, showing the inclination of serrations along the spanwise direction.}\label{fig: figure2}
\end{figure}

Table \ref{table: Design Spec} indicates the range of values for the key geometric parameters of measured barbs found from the barn owl in nature, comparing those with the selected parameter of our generic model, following the data provided in \citep{Bachmann2011}. The definition of the geometric parameters is illustrated in Fig. \ref{fig: figure2}. 
The width is the extension of the major axis of the barb and the thickness is the extension of the minor axis of the barb.
The inclination angle is defined herein between the barb's base and the Z-direction in the X-Z plane (Fig. \ref{fig: figure2}c).
The tilt angle is the angle between the barb's tip and the base in the Y-Z plane (Fig. \ref{fig: figure2}b).
The height and the length of the barb is referred to as H and L as illustrated in Fig. \ref{fig: figure2}.

\begin{figure}[h]
\centering\includegraphics[scale = 0.35]{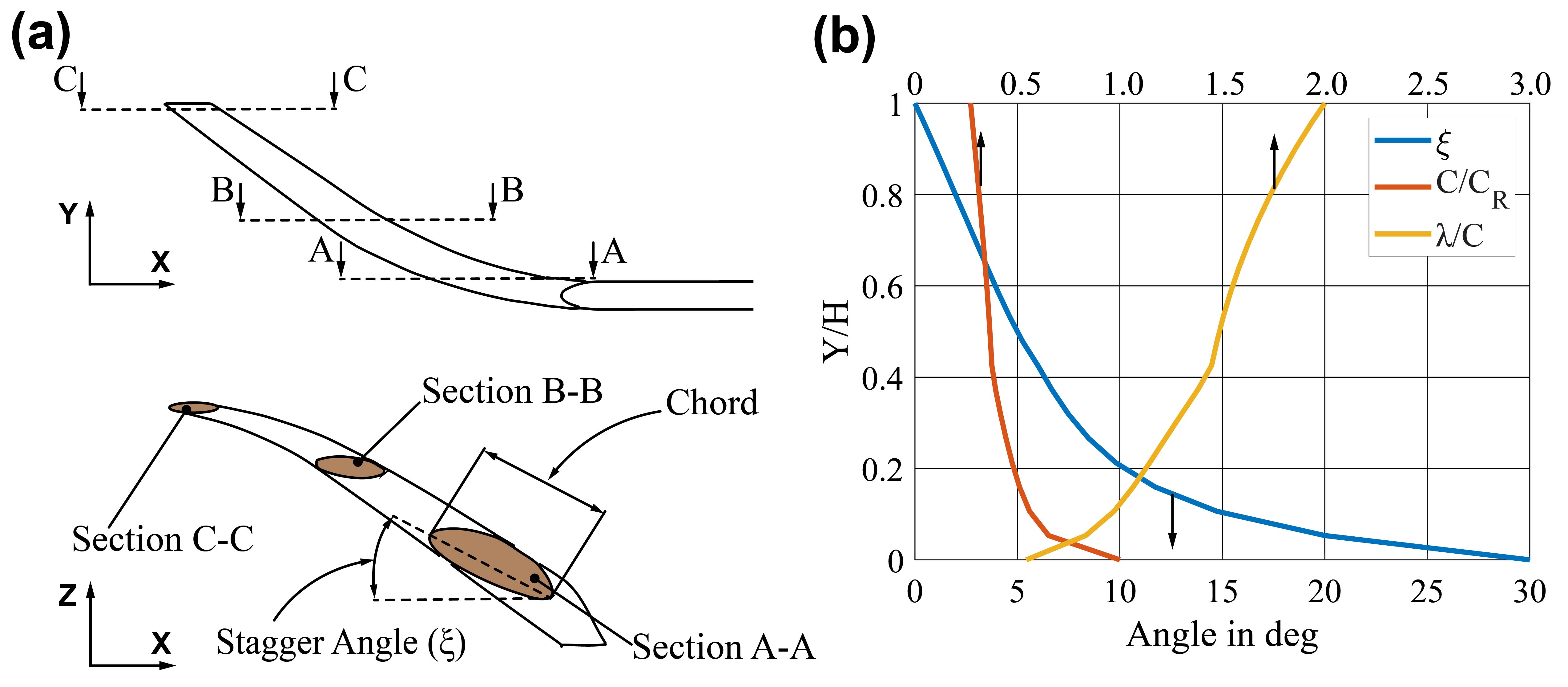}
\caption{ Serration drawings and plots a) Single barb with three sections showing the cross section twist. b) Stagger angle ($\xi$), Normalised chord ($C/C_{Root}$) and spacing to chord ratio ($\lambda$/C) with normalised height of serration}\label{fig: figure3}
\end{figure}

The software SolidWorks (Dassault Syst{\`e}mes, France) was used to design a synthetic barb in the form of a beam with elliptical cross-section (long axis: width, short axis: thickness) and a linear taper from root to tip (root width: 500~$\mu$m, thickness: plate thickness; tip: width: 250~$\mu$m, thickness: 50~$\mu$m) (see Tab. \ref{table: Design Spec}) . 
The length of the initially straight beam was 2250~$\mu$m.
The  elliptical beam was first twisted by 30$^\circ$ (see stagger angle in Fig.\ref{fig: figure3}b, then tilted in the X-Z plane and finally curve-bent in the X-Y plane to reach the desired angles of tilt and inclination given in Tab.\ref{table: Design Spec}. 

In a second step, the root of the beam was then smoothly integrated into the elliptical nose of the flat plate (aspect ratio of about three, thickness of the plate: thickness of the barb at the root) to form the serrated leading edge comb. 
The comb was built as a row of successive barbs with the same spacing (wavelength $\lambda$ = 500~$\mu$m) and size. 
The back, side and top views of the recreated leading edge comb is shown in Fig. \ref{fig: figure2}.
A final qualitative check was done with the geometry of a digitized piece of a 10$^{\text{th}}$ primary feather of an American barn owl (T. furcata pratincola).
The generic model resembled the natural geometry well in all major details of the barb's 3D shape, compare Fig. 1a,b and Fig. 2b,c.

In the following, we interpret the comb as a cascade of blades following the classical nomenclature used in the field of turbomachinery. 
Each blade is represented by one barb and the cascade blade spacing is equal to the comb wavelength.
According to this, we can define the stagger angle as the angle between the chord line of the barb and the axis normal to the leading edge (LE) in the X-Z plane (Fig. \ref{fig: figure3}a) \citep{Dixon}. 
Cross sectional views of individual barbs along the root, middle and tip locations are shown in Fig. \ref{fig: figure3}a. 
The stagger angle is about 30$^\circ$ at the root of the barb and decreases to zero at the barbs' tip. 
Also, the chord decreases along the barbs' height, hence, with same spacing the spacing to chord ratio increases from root towards the tip as shown in Fig. \ref{fig: figure3}b. 

\begin{table}[]
\begin{tabular}{l|l|l}
\textbf{Nomenclature}                          & \textbf{Barn owl data} & \textbf{Idealized model}  \\ \hline
\\
Length ($\mu$m)                   & 1823 -- 2716                     & 1840                    
\\
Wavelength ($\mu$m)                   & 490 -- 670                    & 500    
\\
Width ($\mu$m) @ tip               & 157 -- 215                     & 250                   
\\
Width ($\mu$m) @ root              & 528 -- 652                     & 500                    
\\
\\
Thickness ($\mu$m) @ tip         & 46.9 -- 53.9                     & 50
\\
Thickness ($\mu$m) @ root      & 82 -- 87.2                        &  = plate thickness                   
\\
\\
Tilt Angle ($^\circ$)           & 35.3 -- 36.7                      & 37.5                  
\\

Average Inclination Angle ($^\circ$)   & 50                          & 55.8     
        \\
Angle LE / flight path ($^\circ$)     & 106 -- 138                    & 90 -- 110 \\
                                                         
\end{tabular}
\caption{Dimensions and key geometric parameters of the idealised modeled barb element, leaned upon measurements on barn owls presented by \citet{Bachmann2011}.}\label{table: Design Spec}
\end{table}

\subsection{Numerical Flow Simulations}
\begin{figure}[t] 
\centering\includegraphics[scale = 0.13]{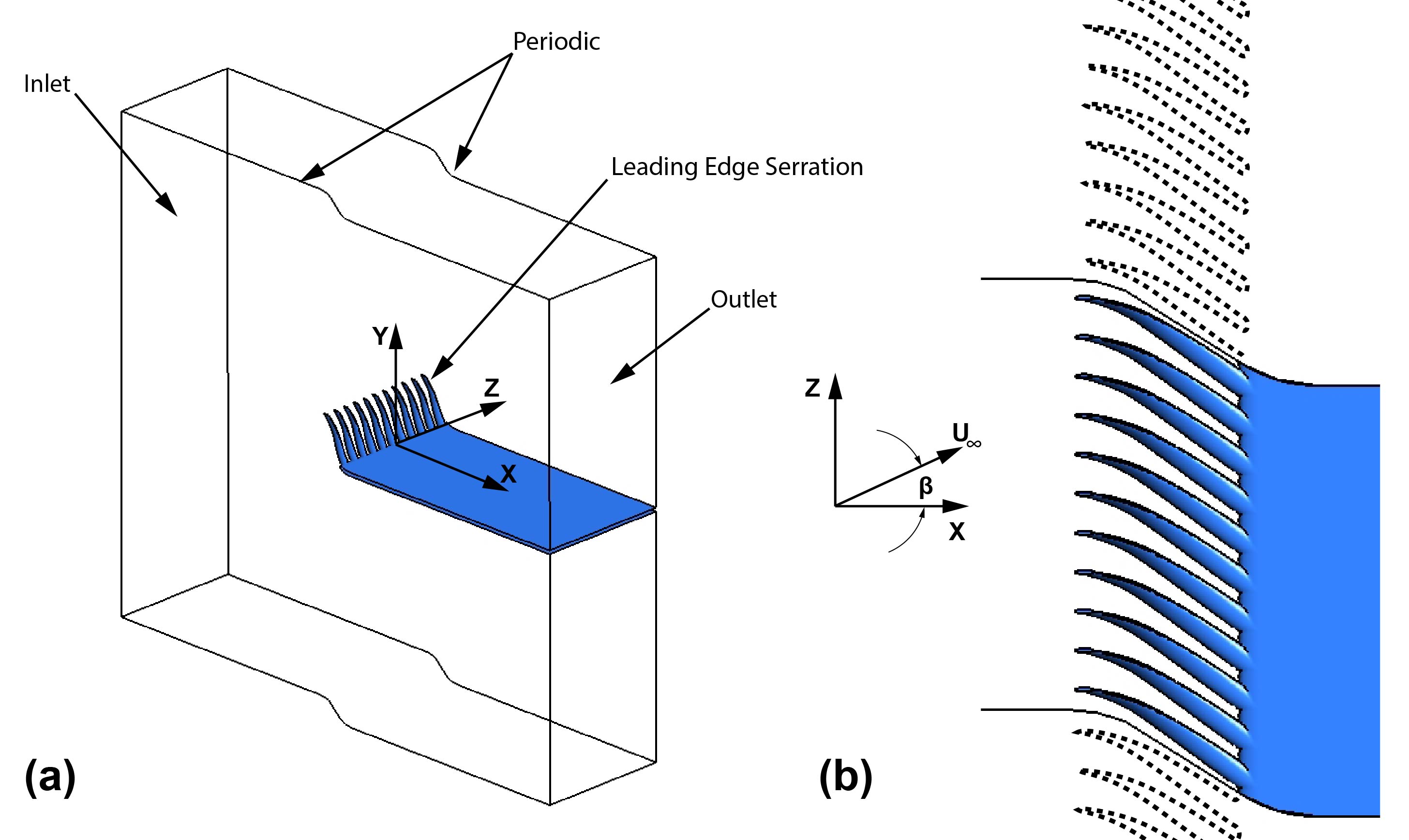}
\caption{Sketches of the CFD domain and the flow configuration with respect to the comb. (a) Isometric view of the CFD domain with periodic conditions in Z-direction. Leading edge serrations attached with the flat plate is shown in blue colour surface (b) Enlarged view of leading edge serration in the X-Z plane showing the direction of the inlet flow velocity vector ($U_\infty$) at an angle ($\beta$) (sweep angle) with X-axis. (Hidden lines of the serration are indicating the periodic boundary condition)}\label{fig: ExpArr_CFD}
\end{figure}

American barn owls have an average wing chord length of $C_W = 0.178$~m \citep{Klan2009} and are supposed to fly with velocities of $U_\infty$= 2.5~m/s to 7~m/s \citep{Bachmann2012}, a number derived from data on European barn owls \citep{Mebs2000}.
At these velocities the Reynolds number $Re_{Wing}$, defined with the wing chord $C_W$, ranges between 30,000 and 100,000, if air temperatures are between 10$^\circ$C and 20$^\circ$C. 
All the simulations and the flow visualisation in the work refer to an average flight speed of 5m/s, which lies within the specified flight-velocity range. For the corresponding  $Re_{wing}$ of 60,000 the boundary layer is in the transitional regime to turbulence, where growing instabilities have an important contribution on noise production. 
Therefore, any possible means to manipulate the flow at or near the leading edge to delay transition may have consequences on the overall flow and acoustic characteristics of the whole wing. 
For our studies, we consider the situation of the animal in gliding flight at constant speed within an otherwise quiescent environment. Therefore, we can chose steady in-flow conditions. 
For the first 10 percent chord of the wing including the barbs on the leading edge, the flow is expected to remain laminar and stationary.
As the barbs have a tiny filamentous shape with a diameter of only few tenth of micron, the local Reynolds-number (built with the chord of the barb) falls around 50, which is small enough that no vortex shedding will occur, see the work of \citep{Paul} for elliptic cylinders.
These conditions pave the way to use a steady-state flow solver in Computational Fluid Dynamics (CFD) to investigate the flow behind the serrations.  Numerical simulations were carried out using ANSYS-Fluent 19.0. 
The wing-fixed coordinate system as defined in §\ref{sec: Method-Coordinates} is used to analyze the data. 
The computational domain extends six serration length upstream and downstream along the X-axis, from the leading edge of the flat plate where the serrations were attached.
Similarly, the domain length in wall-normal direction (Y-axis) extends five serration lengths in either direction and the spanwise direction (Z-axis) has a length which accommodates 11 serrations as shown in Fig.\ref{fig: ExpArr_CFD}a.
The domain is meshed with tetrahedral elements with inflation layers near the wall, furthermore, the mesh was refined near the serrations to capture the flow gradients accurately, resulting in a mesh size with about 19 million cells.
Computations were performed with a steady-state solver and the $k-\omega$ model for solving the RANS turbulence equations. 
At the inlet  a constant free stream velocity ($U_\infty$) is assumed. 
The direction of this velocity vector relative to the coordinate system of the wing and the leading edge indicates whether the flow is facing a swept wing or not.
Zero sweep means that the leading edge is aligned with the outboard directed spanwise axis of the flying body and the inflow velocity vector is parallel to the chord-wise axis of the wing ($\beta = 0$ relative to the X-axis in the X-Z plane) as shown in Fig. \ref{fig: ExpArr_CFD}b. 
To simulate the sweep effect of the wing, the angle $\beta$ was varied from -10 degree (forward swept wing) to +20 degree (backward swept wing). 
Constant pressure was assumed at the outlet and periodic boundary conditions were given at the lateral sides, which results in infinite repetitions of the serrations (neglecting end effects).

\subsection{Flow Visualization}
\begin{figure}[t] 
\centering\includegraphics[scale = 0.95]{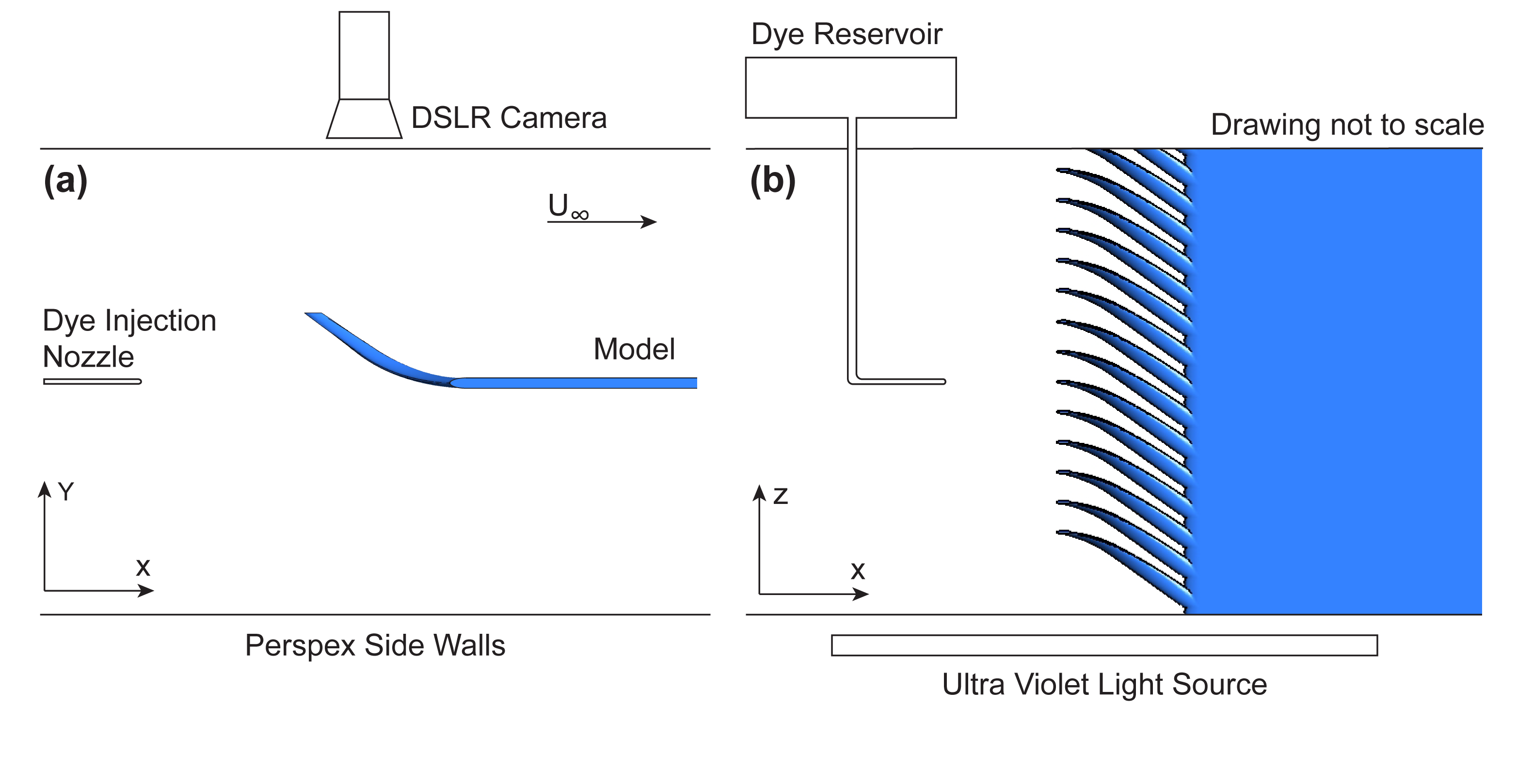}
\caption{Sketches of the experimental set-up for the dye flow visualizations carried out in the CHB Water Tunnel at City, University of London. (a) plan view of the set-up in the horizontal cross-section. (b) Side view on the vertically mounted flat plate.}\label{fig: ExpArr}
\end{figure}

For the experimental flow studies, the model of the flat plate with the leading-edge comb was 3D printed with a 20:1 upscaling factor (Stratasys OBJET 30 PRO printer with a print accuracy of 30 microns, material Veroblack).
Fabrication of the serrations in original size was discarded after tests of different micro-manufacturing methods showed extreme difficulties to reproduce the shape of the barbs in good quality with the current available technology.
With the given up-scaling factor, the method of dynamic similitude in fluid mechanics \citep{Durst} provides the corresponding boundary conditions for flow visualization studies in a wind- or water-tunnel, the latter being more suited herein. 
Dye flow visualisations were carried out in the CHB Water tunnel facility at City, University of London. 
The tunnel is a closed loop, open surface tunnel which operates horizontally with a 0.4~m wide, 0.5~m deep and 1.2~m long test section. 
According to the laws of similitude, the freestream velocity of the water was set to 3.3 cm/s, corresponding to the situation of 5 m/s in air with the serration in original scale.
The leading edge of the up-scaled model was placed vertically in the tunnel 0.4~m downstream of the entrance of the test section, extending from the floor of the tunnel up  to the free water-surface (Fig. \ref{fig: ExpArr}). 
This situation reproduces the flow along the flat plate with zero sweep of the leading edge.   
Fluorescent dye was injected through a tiny needle (1~mm inner diameter, 1.6~mm outer diameter) which was placed upstream of the model (Fig. \ref{fig: ExpArr}b). 
Care was taken to control the dye exit velocity the same as the bulk fluid flow. 
This is crucial to avoid instabilities of the fine dye streakline  ultimately compromising the result \citep{Merzkirch1987}. 
An ultra-violet (UV) lamp was placed underneath the perspex floor of the test section to enhance the contrast of the fluorescent dye against the background. 
A NIKON D5100 DSLR camera was used to capture the resulting flow visualization (Fig. \ref{fig: ExpArr}a). 
The camera was mounted on a tripod and was situated parallel to the surface of the model, to observe the evolution of the dye filament on the surface of the model. 
Due to the low light level, a long exposure (20 seconds) image was taken with the lens aperture set to f/10.
Such a long-time exposure is allowed as the flow pattern remained stationary, indicating a steady flow situation. 
The images were then subsequently enhanced using `Adobe Photoshop' to provide better clarity.  

\section{Results}
In the following we present both experimental and simulation data on a new hypothesis on the function of the serrated comb of the leading edge of the owl wing. 
The new hypothesis states that the 3D curvature of the serrations cause a change in the direction of the flow inboards towards the owl’s body (called ``flow turning" in the following), in this way counteracting the outboards directed cross-span flow induced by the backward sweep of the wing. 
We first show the basic predictions of our model and the validation of these predictions by experiments in a water tank. 
In a second part we examine the properties of the flow turning in more detail.\\

\subsection{Basic results of experiments and CFD simulations}
\begin{figure}
\begin{subfigure}[t]{.45\textwidth}
  \centering
  \includegraphics[width=1\linewidth]{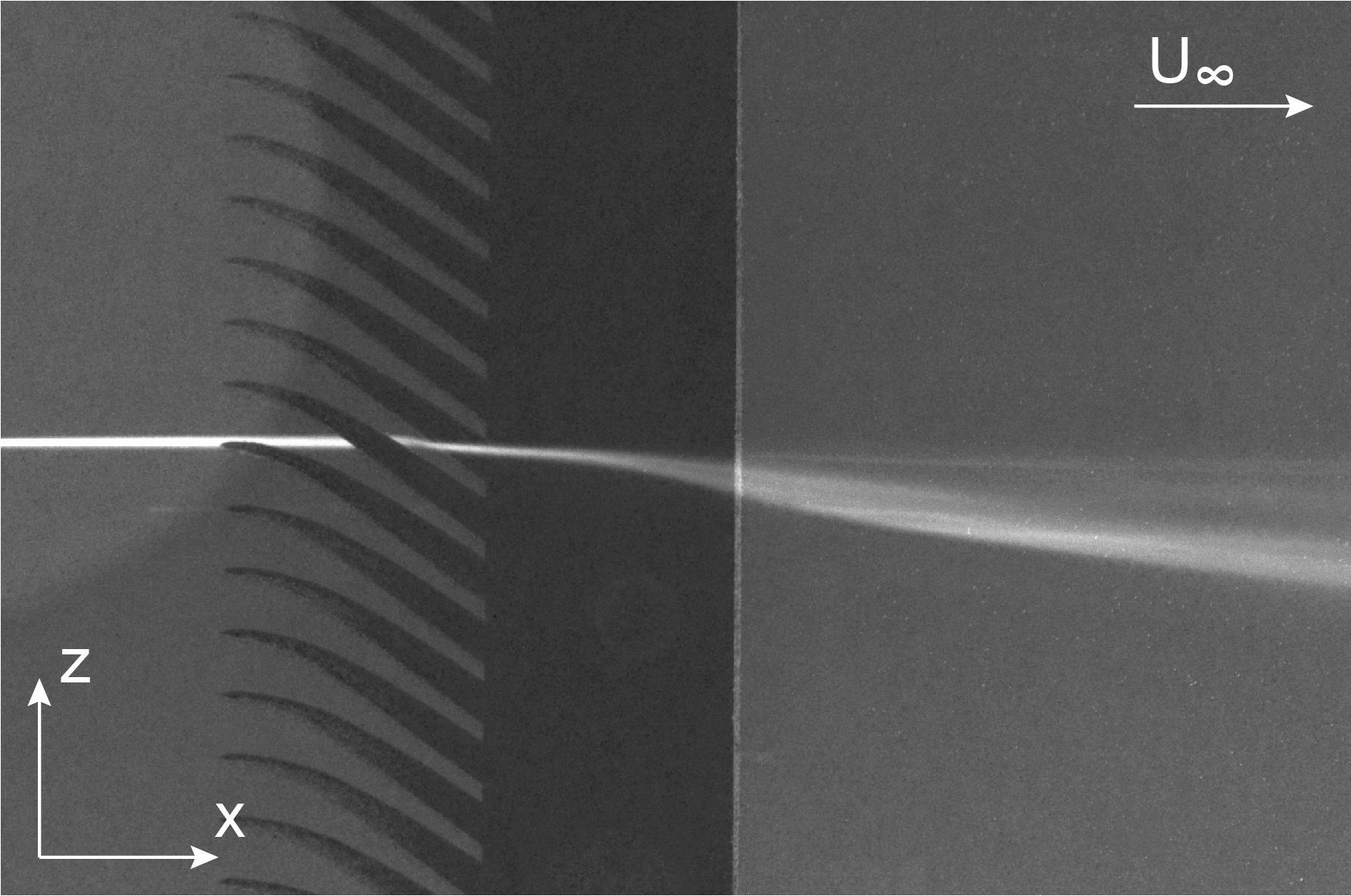} 
\caption{Long-time exposure image of the dye flow visualisation, illuminated under ultra violet light (image has been contrast-enhanced for better clarity).}\label{fig: FlowVis}
\end{subfigure} \hspace{3mm}
\begin{subfigure}[t]{.45\textwidth}
  \centering
  \includegraphics[width=1\linewidth]{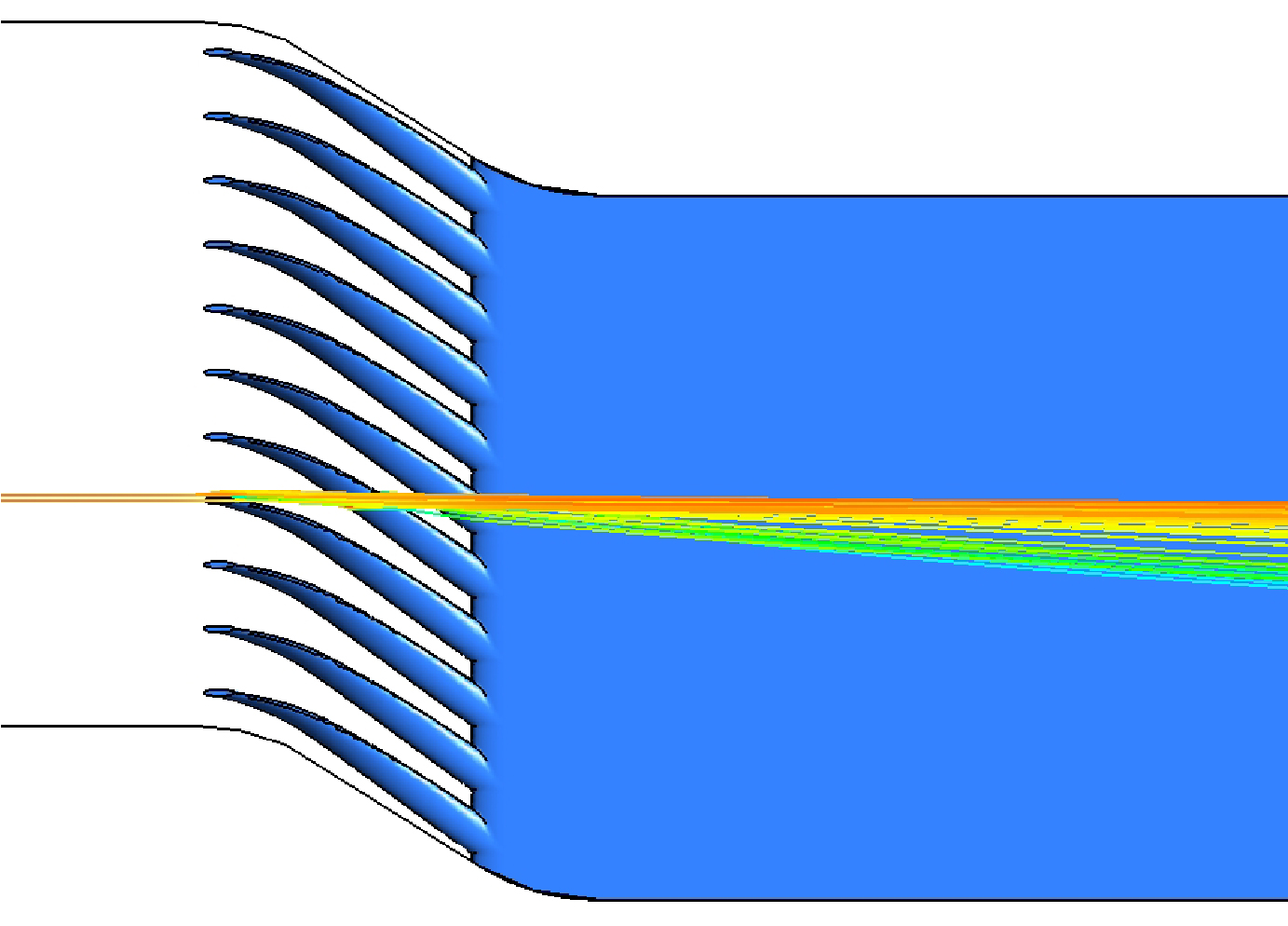}  
  \caption{Top view on streamlines  with different starting points along the wall-normal axis in color (green: near-wall to red: tip of the serrations, CFD simulation at $\beta$ = 0\si{\degree}).}
  \label{fig: CFD Steamlines}
\end{subfigure}

\begin{subfigure}{\textwidth}
  \centering
  \includegraphics[scale = 1]{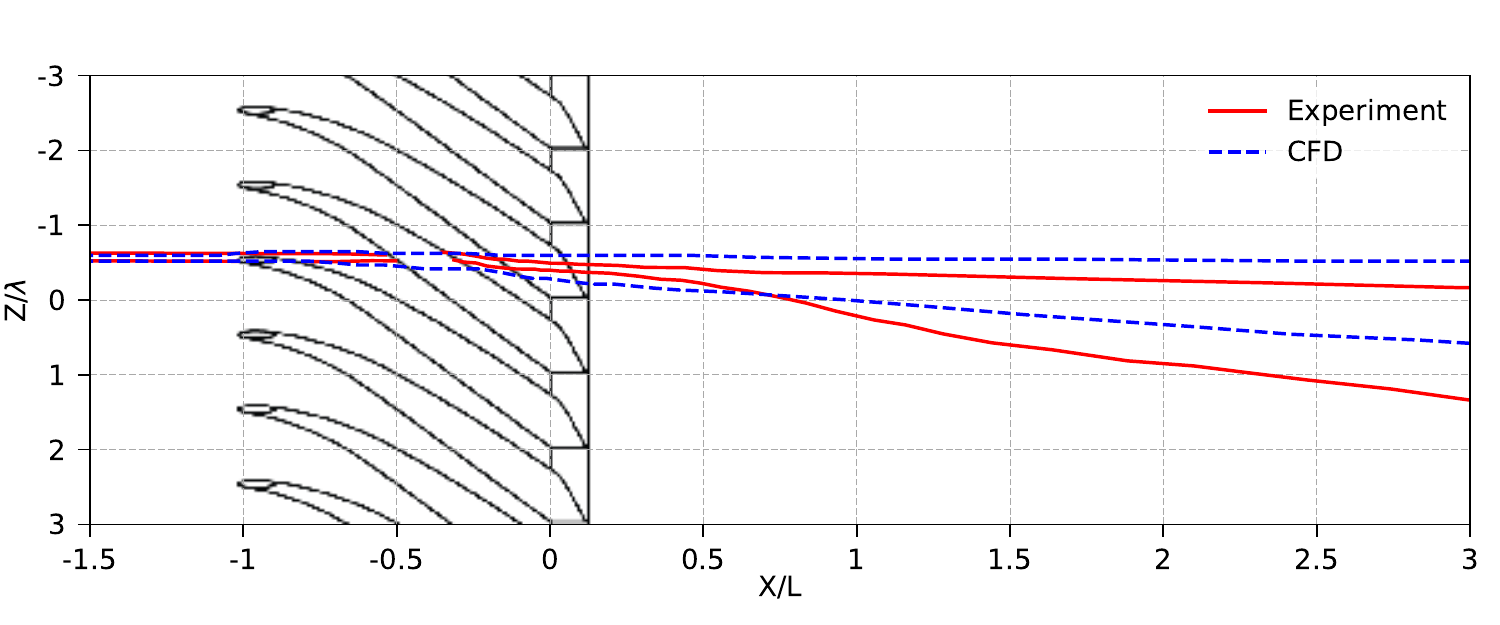}
  \caption{Range of the most-extreme turning streamline relative to the streamline at the tip.  From the CFD simulation and the dye trace from the water tunnel experiment}
  \label{fig: CFDvsFlowVis}
\end{subfigure}
\caption{Comparison of flow visualisation and CFD results. }
\label{fig: Comparison}
\end{figure}

\begin{figure}[h]
\centering\includegraphics[scale = 0.12]{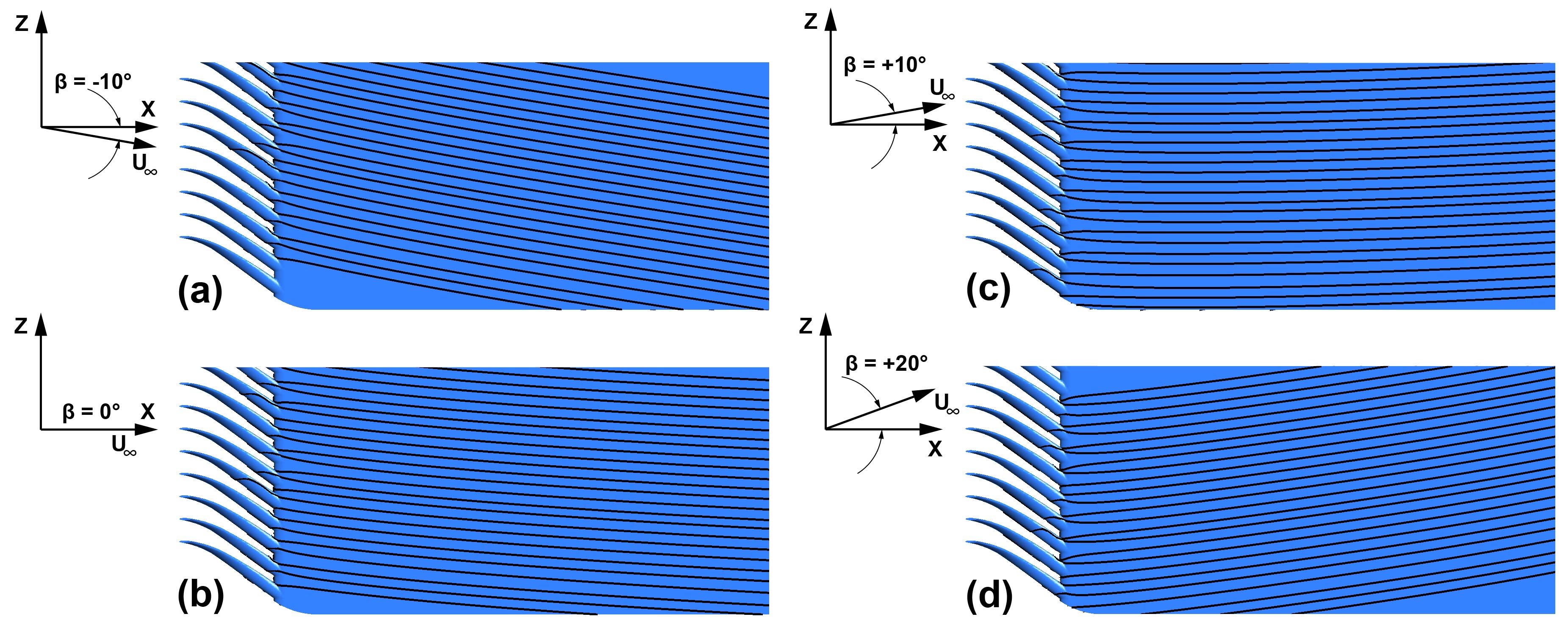}
\caption{Surface streamlines from CFD simulations. (a) Negative sweep angle $\beta$ = -10\si{\degree}. (b) Zero Sweep angle $\beta$ = 0\si{\degree}. Positive sweep angle (c) $\beta$=+10\si{\degree}. (d) $\beta$=+20\si{\degree}}\label{fig: figure7}
\end{figure}

Figure 6 shows the streamlines (Fig.\ref{fig: FlowVis}  experiment, Fig.\ref{fig: CFD Steamlines} computed from the steady state CFD simulation) when started upstream from the serrations to downstream of them, first analyzed for the situation of zero sweep. The flow situation in the water tunnel with dye flow visualization shows a white coloured thick streamline upstream of the serrations in direction parallel to the X-axis. Once the water passes the serration, a flow turning effect can be seen as the streamline points downwards  at a certain angle in negative Z-direction (inboards). Furthermore, the visualization shows that the flow remains laminar and steady. This justifies our decision to use a steady-state flow solver. The near-surface streamlines generated from the CFD results look very similar to the experimental result (Fig. \ref{fig: CFD Steamlines}). The different colours indicate different streamlines started at the same X,Z-location but at varying wall-normal distances `Y' to the flat plate. Near the wall (blue to green colours), the flow turning is maximum, it reduces with increasing distance from the plate and disappears completely at the serration tip (red colour). This indicates an induced cross flow near the wall. We interpret this data such that the 3D curved shape of the serrations cause this change in flow direction, because in a plate without serrations or a plate with symmetric planar serrations such a change in flow direction is not expected to occur. In Fig. \ref{fig: CFDvsFlowVis} the envelope of the flow turning effect is given by the two extreme streamlines, the one with zero and the one with maximum turning, respectively, for both the CFD and the flow visualization in Fig.\ref{fig: CFDvsFlowVis}. Since the result from the flow visualisation and the CFD are in good agreement, further results from CFD simulations can be accepted with confidence. 
Fig.\ref{fig: figure7} shows the near-surface streamlines (along the first cell away from the wall of the numerical mesh) on the flat plate surface for various inlet flow angles in the X-Z plane. In Fig.\ref{fig: figure7}b the inlet flow is aligned with X-axis (zero sweep) and once the fluid passes through the serration the flow is turned towards inboard direction as already explained above. The same trend of flow turning is observed also for increasing backward sweep (angle $\beta$ = 10\si{\degree} Fig. \ref{fig: figure7} c and 20\si{\degree} Fig. \ref{fig: figure7}d). Altogether, this data proves that the serrations work as a cascade of guide vanes or finlets which turn the flow in the boundary layer in the opposite direction of the normally observed cross-span flow in a coherent manner along the span.

\subsection{Detailed examination of the flow turning}
\begin{figure}[ht]
\centering\includegraphics[scale = 0.5]{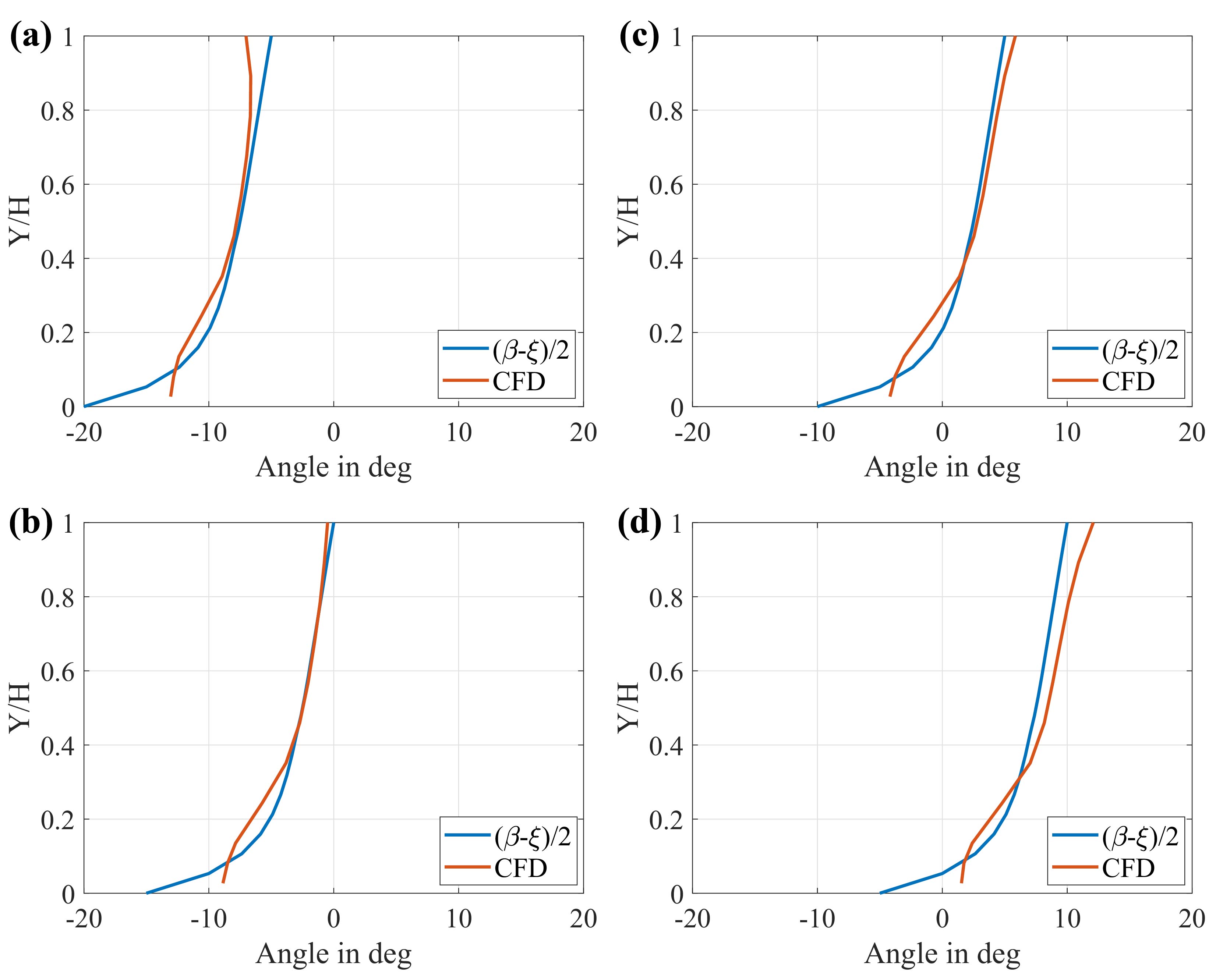}
\caption{Wall-normal variation of turning angle behind serrations at X/L=0 for different sweep angles from CFD results and analytical formula. (a) $\beta$ = -10\si{\degree}. (b) $\beta$ = 0\si{\degree}. (c) $\beta$ = 10\si{\degree}. (d) $\beta$ = 20\si{\degree}.}\label{fig: figure8}
\end{figure}

Further information is gained from the flow turning angle just behind the serrations shown in Fig.\ref{fig: figure8} for various inlet flow angles. 
As the chord and the stagger angle are largest at the root of the barbs (Fig. \ref{fig: figure3}b), it is obvious that the flow turning is more pronounced near their root, while it reduces when moving towards the tip. 
We again take help from the similarity to stationary guide-vanes and approximated the flow turning angle as proportional to the difference between inlet flow angle ($\beta$) and the stagger angle ($\xi$). The correlation of the turning angle equal to ($\beta-\xi$)/2 is based on the classical exit flow angle formula used for cascade blades $\xi$ \citep{Dixon}.
For cases with inlet flow angle of $\beta$ = 0 and +10 degrees the correlation is reasonably good (Fig.\ref{fig: figure8}b and Fig.\ref{fig: figure8}c), even for larger $\beta$ = +20 degrees the trend is captured quite well (Fig.\ref{fig: figure8}d). 
The observed correlation captures the overall trend based on considerations for classical 2D guide vanes, indicating that even though the serrations have a 3D curved shape,  the main factors in defining the flow turning is mostly determined by the dimensional variation of the chord and the stagger angle.

\begin{figure}[h]
\centering\includegraphics[scale = 0.5]{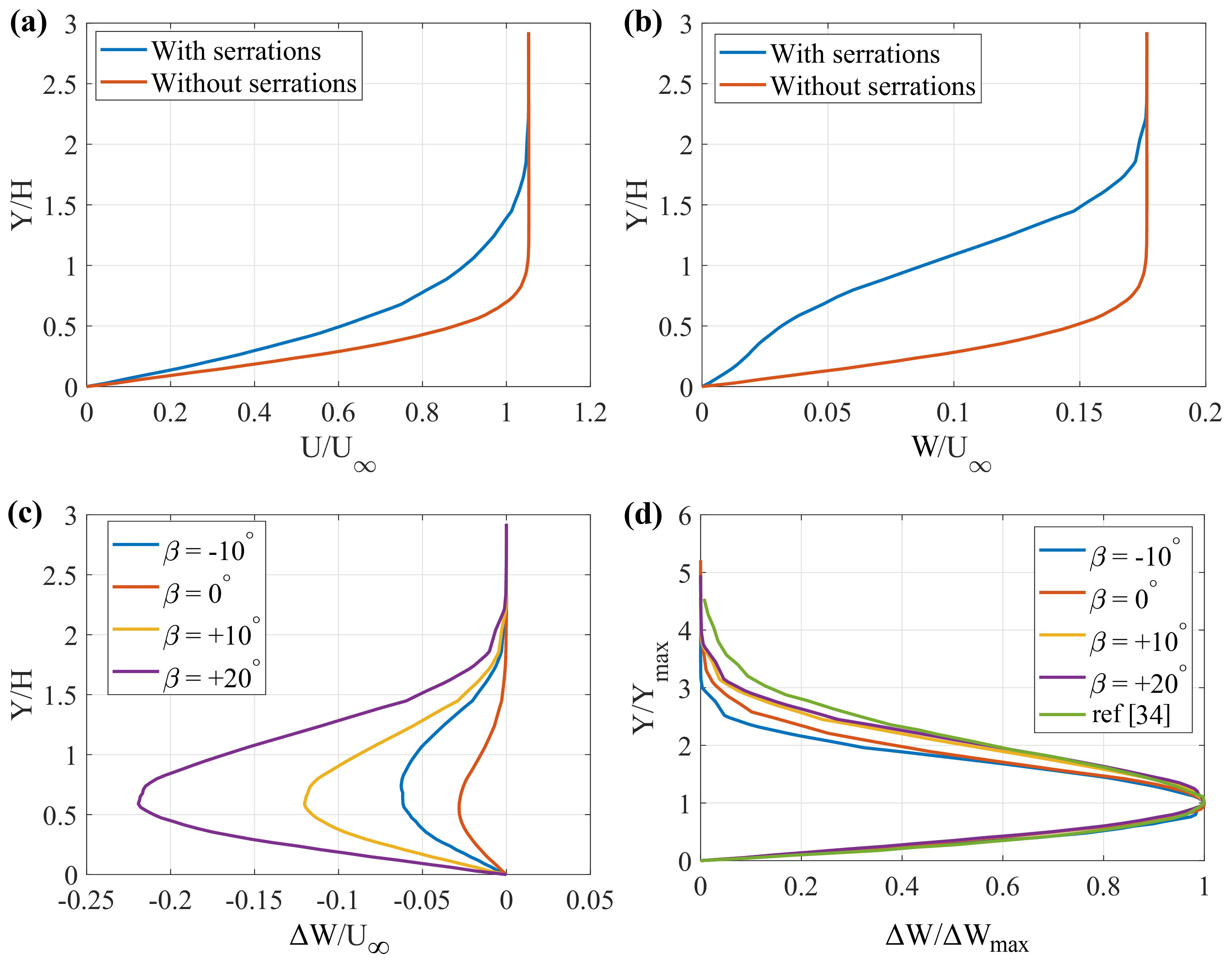}
\caption{Velocity profiles from CFD simulations at X/L = 6 downstream of the leading edge. (a) Chordwise velocity for $\beta$ = +10\si{\degree}. (b) Spanwise velocity for $\beta$ = +10\si{\degree}. Net-effect of cross-flow profile (c) For all sweep angles. (d) Normalised cross-flow velocity profile with comparison to \citet{Ustinov2018}}\label{fig: figure9}
\end{figure}

Note, that the flow turning effect induced at the plane of the serrations is affecting the direction of the streamlines even far downstream the chord until at the downstream end of the simulation domain (Fig. \ref{fig: CFDvsFlowVis}), see also the flow visualisation experiment. Therefore the serration have a far-reaching effect on the boundary layer flow down the chord. 
To show that, we compared simulations for the plain plate with those having attached the leading-edge comb under otherwise identical boundary conditions.
Normalised chordwise and spanwise velocity profiles at the outlet section at X/L=6 for a sweep angle of 10 degrees are shown in Fig.\ref{fig: figure9}a and Fig.\ref{fig: figure9}b. 
With serrations, the chordwise velocity profile shows a larger deficit than without serrations (Fig. \ref{fig: figure9}a), which leads to an increase of the displacement ($\delta^*$) and momentum thickness ($\theta$) to twice the value without serrations (flat plate). 
However, the shape factor ($H = \delta^*/\theta$) remains around 2.4, suggesting that the serrations are not acting as a flow tripping device (this is when the shape factor gets beyond 3.5). 
The spanwise velocity profile for the plain plate (without serrations) resembles the one in chordwise direction (Fig. \ref{fig: figure9}b). 
However, adding the leading-edge comb leads to a dramatic decrease of the spanwise flow inside the boundary layer region with further reach into the free-stream. 
For a better illustration of the net-effect induced by adding the leading-edge comb, we plot the difference of the spanwise velocity profile ($\Delta$W) defined as $W_{wi}-W_{wo}$ for all the cases considered here (wi - with serrations, wo - without serrations).This resultant velocity profile increases from zero to a maximum value within half the height of the barb and then it monotonically decays to minimal value at a height which is more than twice the height of the barb. 
Hence, this profile strongly resembles that of a wall jet, which counter-acts the sweep-induced spanwise flow in the plain plate (Fig. \ref{fig: figure9}c). 
The peak values in $\Delta$W are reached at about half the serration height for all flow angles. Furthermore, the magnitude of the peaks increase with increasing sweep angle. 
These results show also a significant flow turning effect for the negative sweep angle ($\beta = -10^\circ$), which was not clearly recognizable from the illustration of the surface streamlines (Fig. \ref{fig: figure7}a).

When all the $\Delta$W profiles are normalised with respect to their corresponding maximum and the coordinates are scaled with respect to the position of maximum velocity, the profiles nearly collapse (Fig.\ref{fig: figure9}d). 
The data well resembles the spanwise velocity profile used in the theoretical work from \citet{Ustinov2018} that was effective in counter-acting the cross-wise instabilities in swept wing flows. 
  
\section{Discussion and Conclusions}
We showed that serrations at the leading edge of an owl inspired model induce an inboard directed flow that is in opposite direction to the cross-span flow induced by the backward sweep of the wing. 
In the following we shall first discuss these data with respect to the existing literature, arguing about some methodological considerations and then speculating about its consequences for owl flight and flight in general.

\subsection{Comparison with other work}
To the best of our knowledge, no study has directly addressed how the sweep angle influences the flow in nature-inspired serrated wings. The work most important to our new data and hypothesis is that by \citet{Ustinov2018}. 
The near overlap of the curves in Fig.\ref{fig: figure9}d shows that the serrations reproduce the effect envisioned by \citet{Ustinov2018}.
These authors discussed this effect as to counter-acting the cross-wise flow in swept wing and thereby attenuating the crossflow instabilities, a negative feature of backward swept wing aerodynamics. 
The work of these authors is based on a theoretical consideration of micro-perforation or winglets on the surface of a wing, which are arranged in a way that they produce a spanwise flow in the boundary layer opposite in direction to the cross-span flow induced by the sweep-effect. 
With this configration, \citet{Ustinov2018} observed a wall-jet like flow profile in spanwise direction that is similar in shape and relative magnitude to our net-effect result. 
Therefore, the 3D curved serrations of the barn owl wing could be thought of as a leading-edge laminar flow control device which counteract the cross flow instabilities in swept wing aerodynamics.

As we could show here, the serrations of the Owl wing are not comparable to classical vortex generators, which was speculated so far in previous work \citep{geyer2020, Hertel1963}.
These vortex generators are used traditionally to control the flow separation on the suction side of the airfoils \citep{linjohn}.
They produce strong streamwise vortices to mix the fluid flow via the lift-up effect, thus increasing streamwise momentum near the wall. 
In comparison, our study found that the serrations studied herein, behave similar to 3D curved cascade blades which turn the flow to a certain degree depending on the spacing to chord ratio and the blade angle (stagger angle).
Hence, near the root of the serrations the spacing to chord ratio is low and the stagger angle is high to guide the flow to turn at relatively high angles when compared with the tip. \citet{Kroeger1972} hinted on the cascading effect of the leading edge serrations. However, they stated that the serrations push the flow behind the leading edge towards the outboard region of the owl wing, which is opposite to our observation. Note, that their statement resulted from tuft flow visualisation where the length of the tufts was greater than 4~mm. Therefore, the tuft motion will be the result of an integration all over the complete boundary layer thickness and part of the external flow. Since the height of the serrations is less than 2~mm, they probably could not see our results because of this integration effect. In addition, any method of flow visualization or flow measurement must ensure to get data very close to the wall as provided herein. This is where we benefit from the testing of an enlarged model in a water tunnel, fulfilling the rules of fluid mechanical similitude. 

A vague indication of flow turning may be found in the results from \citet{wei2020}, although not mentioned therein. It seems from their Fig. 10b in \citet{wei2020}) that the hook-like serrations changed the direction of flow. However, since the graph is cut downstream at about 0.5 of serration length, it is difficult to infer a concluding answer on any flow turning.

\subsection{Methodological considerations}
It is obvious from live recordings of the gliding flight of owls that the leading edge in the region of serrations is swept backward  \citep{Kroeger1972, Durston2019}, an aspect which has so far not found attention in the discussion of the function of the serrations. We observed a flow turning effect induced by the 3D curved serrations, which counter-acts the crossflow induced in backward-swept wing. In this respect it seems important that we have carefully rebuilt the natural shape of the serrations, characterized by twisting and tilting and taper, which Bachmann and Wagner \citep{Bachmann2011} called a first order approach and not used the zero order approach, i.e. use simply-shaped, often symmetric serrations as is done in most studies\citep{geyer2020, Rao2017,ikeda2018, geyer2017-LEC}. 
The focus of the study was to demonstrate the basics of the novel turning effect. A good correlation was found between the observed turning angle and the classical formula for cascade blades, approximated as the summation as inlet flow angle $\beta$ and the stagger angle $\xi$ \citep{Dixon}.

Not all parameters could be assessed in this first study. Further work might unravel the role of the wavelength, as it is obvious that a too large inter-spacing will destroy the homogeneity of the induced crossflow and a too small inter-spacing will cause unnecessary form drag. 
More studies are also necessary to find out how the angle of attack and the Reynolds number influences the flow turning, and how far the laminar hypothesis is valid.

\subsection{Consequences for owl flight}
The consequence of a manipulation on the flow reported in \citet{Ustinov2018} for a swept wing is that it delays transition to turbulence. 
Because of the striking similarity of the effect of the manipulation on the boundary layer profile to the effect we observed, we conclude that the leading-edge comb acts to delay transition on the swept wing of the owl. 
A delay of transition would correspond to a reduction in noise production as the portion on the wing surface where the flow is turbulent is reduced or even completely removed. 
Owl flight is so silent that it is difficult to measure directly (in absolute terms) the noise these birds produce. Only in comparison with other, non-serrated wings, does the noise-reduction of owl flight become clear \citep{neuhaus1973, geyer2020}. Thus, the influence on the air flow as demonstrated here may be critical in nature, where a hunting owl has to remain silent until right before the strike.
Serrations which can help to keep the flow laminar and preventing cross-flow instabilities for typical flight conditions with backward swept wing , therefore, may provide a major advantage for the hunt.

\subsection{Conclusions}
To conclude, we have investigated the effect of a nature-inspired leading edge comb on the flow along a swept flat plate. 
Special focus is laid on the leading-edge comb influence on the backward swept wing in gliding flight, which is known in classical wing aerodynamics to introduce considerable cross-span flow, which suffers instabilities and triggers early transition \citep{serpieri, Radeztsky, Edward}. 
As evidenced in the CFD and the experiments, our model produces a flow turning which is counter-acting the cross-span flow. The magnitude of this effect is proportional to the stagger angle of the local cross-section of the barbs. If the sweep angle is increased, the flow turning becomes more pronounced, suggesting that the owl's leading-edge comb is tailored for attenuating the cross-flow instabilities. Ultimately, this means a laminar flow control with benefit of a quiet flight.

\section*{Acknowledgements}
The position of Professor Christoph Bruecker is co-funded by BAE SYSTEMS and the Royal Academy of Engineering (Research Chair No. RCSRF1617$\backslash$4$\backslash$11, which is gratefully acknowledged. The position of MSc Muthukumar Muthuramalingam was funded by the Deutsche Forschungsgemeinschaft in the DFG project BR~1494/32-1 and  MEng Edward Talboys was funded by the School of Mathematics, Computer Science and Engineering at City, University of London.
Hermann Wagner was supported by RWTH Aachen University. 
We like to thank Matthias Weger, Adrian Klein and Horst Bleckmann for discussion on the owl's leading edge geometry and its relevance to silent owl flight. 

\section*{Author contributions statement}

All authors conceived the experiment(s),  M.M. and E.T. conducted the experiments, all authors analysed the results. Initial draft was prepared by M.M, E.T and C.B. The finalised version was prepared with the contribution from all authors.

\section*{Additional information}

\textbf{Accession codes} (where applicable);\\ \textbf{Financial Competing interests} The authors declare no competing interests. \\ \textbf{Non-Financial Competing interests} The authors declare no competing interests.  
 
\newcommand{\newblock}{}

\bibliographystyle{unsrtnat}
\bibliography{OwlLeadingEdge}

\end{document}